\documentclass[aps,twocolumn,superscriptaddress,showpacs,showkeys]{revtex4}
\usepackage{graphics,graphicx,dcolumn,bm,fleqn,epic,eepic,float}
\usepackage{amssymb,amsmath,multirow,rotate,color,float,times}

\definecolor{blue}{rgb}{0,0,1}
\definecolor{red}{rgb}{1,0,0}

\begin{document}
\title{Spreading gossip in social networks}

\date{\today}

\author{Pedro G.~Lind}
\affiliation{Institute for Computational Physics, 
             Universit\"at Stuttgart, Pfaffenwaldring 27, 
             D-70569 Stuttgart, Germany}
\affiliation{Centro de F\'{\i}sica Te\'orica e Computacional, 
             Av.~Prof.~Gama Pinto 2,
             1649-003 Lisbon, Portugal}
\author{Luciano R.~da Silva}
\affiliation{Departamento de F\'{\i}sica Te\'orica e Experimental, 
             Univ.~Federal do Rio Grande do Norte,
             Campus Universit\'ario, 59072-970 Natal-RN, Brazil}
\author{Jos\'e S.~Andrade Jr.}
\affiliation{Departamento de F\'{\i}sica, Universidade Federal do Cear\'a,
             60451-970 Fortaleza, Brazil}
\author{Hans J.~Herrmann}
\affiliation{Departamento de F\'{\i}sica, Universidade Federal do Cear\'a,
             60451-970 Fortaleza, Brazil}
\affiliation{Computational Physics,
             IfB, HIF E12, ETH H\"onggerberg, CH-8093 Z\"urich, Switzerland}

\begin{abstract}
We study a simple model of information propagation in social networks,
where two quantities are introduced:
the spread factor, which measures the 
average maximal fraction of neighbors of a given node that
interchange information among each other, and the spreading time needed for 
the information to reach such fraction of nodes.
When the information refers to a particular node at which both 
quantities are measured, the model can be taken as a model for gossip 
propagation.
In this context, we apply the model to real empirical networks of 
social acquaintances and compare the underlying spreading dynamics with 
different types of scale-free and small-world networks.
We find that the number of friendship connections strongly influences 
the probability of being gossiped.
Finally, we discuss how the spread factor is able to be applied to
other situations.
\end{abstract}


\maketitle


\section{Introduction and model}
\label{sec:intro}

In every-days life probably everyone has already experienced the
annoying situation of telling some personal secret to some friend
and ending with a naive ``please, do not tell that to anyone, ok?''
and after short time all our friends suddenly know the secret.
What happened? Is this common phenomenon a consequence of a natural
instinct that friends have to conspire and slander against each other?
Or is this a phenomenon which can hardly be avoid by human trust and
respect being closely related to the net of acquaintances that
people naturally tend to form?
\begin{figure}[htb]
\begin{center}
\includegraphics*[width=8.0cm,angle=0]{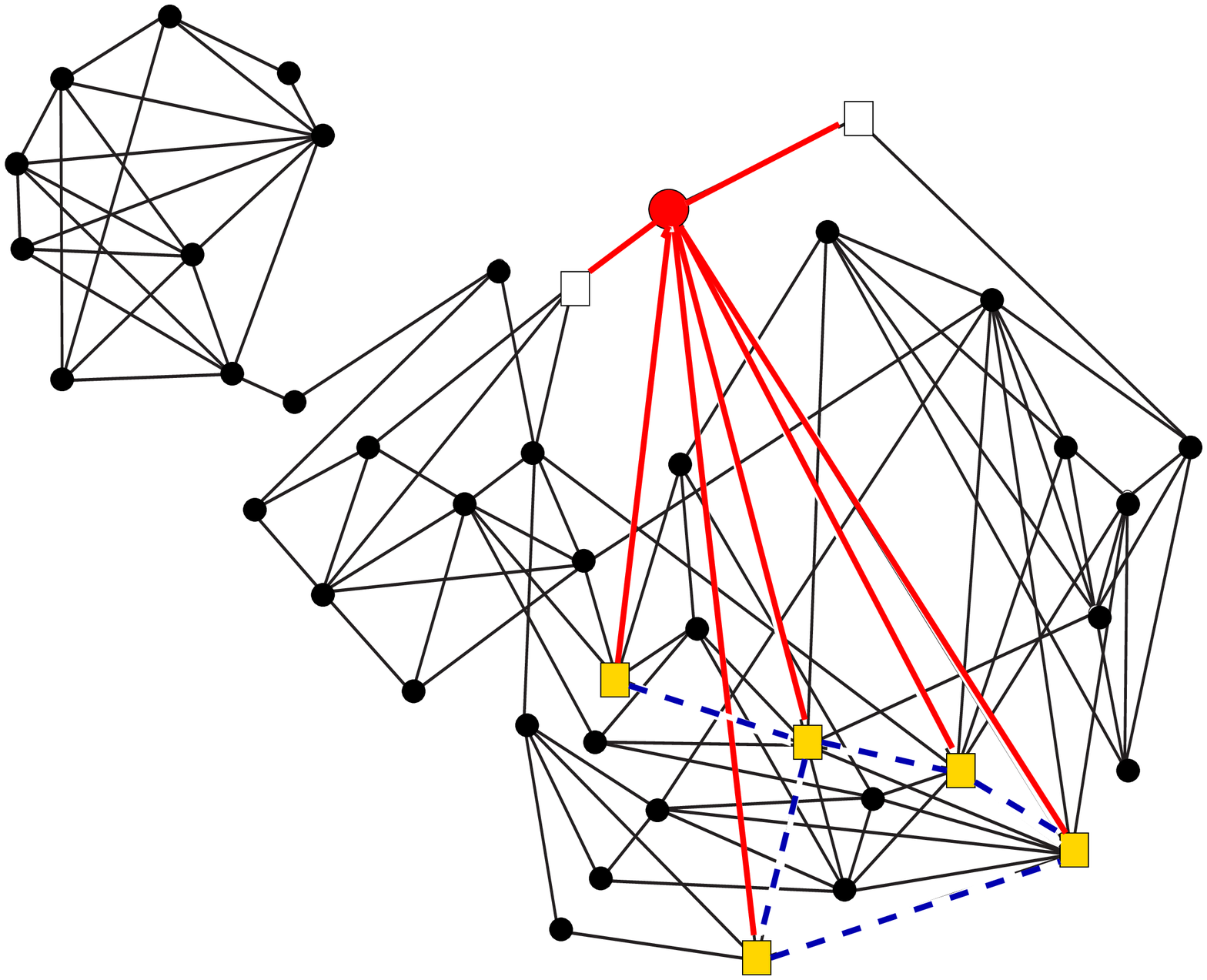}
\end{center}
\caption{\protect 
     (Color online)
     Spreading of information about a target-node shown as the
     grey (red) open 
     circle on part of a real school friendship network~\cite{schools}.
     If the spreading starts from one of the white squared neighbors, no 
     propagation occurs ($f=0$). If instead, one of the grey (yellow) squared 
     neighbors starts the spreading, in $\tau=3$ time-steps, five neighbors 
     will know it, giving $f=5/7$. 
     The information spreads over the dashed (blue) lines.
     The information can be seen as a gossip about the target-node or the 
     victim (see text).
     Note that the clustering coefficient of the victim has a
     different value, namely $C=10/42$.} 
\label{fig1}
\end{figure}

Such kind of questions can be easily addressed by representing the social
system, composed by individuals and the interactions among them, as a network,
i.e., as a collection of nodes and links.
While networks have been widely used by physicists to study e.g.~porous 
media~\cite{hans05} or a system of interacting 
spins~\cite{sanchez02,krapivsky03,mobilia03}, they can also be
used to study social systems. 
Social networks have helped to further understand the structure and 
evolution of social systems, where people and their acquaintances
are represented by the nodes and links of the network respectively.
In particular, propagation of information in social systems is easily 
reproduced in such networks and has been addressed
in recent physical literature~\cite{boccaletti05,shao03,cycles}
due to its importance in epidemiology~\cite{dodds04}, where information 
is related to the contagious of diseases, to understand social influence, 
beliefs and extremism~\cite{castellano00,pluchino05,galam05,he04}, 
to understand the evolution of financial markets~\cite{eguiluz00}, to 
study econophysical networks underlying e.g., electrical supply systems or 
road webs among airports or cities. 
Here we put emphasizes on how far the information can spread when 
particular constraints, of interest for social systems, are taken into 
account.

The way information spreads over the network depends on its content.
A rumour or an opinion concerning some topic which is
not directly connected to the social network structure (political opinion,
etc) can be of interest to any of the neighbors of a certain node,
regardless their topological features.
However, as opposed to rumors, a gossip always targets the details about
the behavior or private life of a specific person, i.e., of a specific node.
This node will be called henceforth the target-node or the victim.
Therefore, due to this particular content, it is reasonable to assume as a 
first approach that the information spreads only over people directly 
connected to the victim.

A simple model recently introduced~\cite{us} for such kind of information 
spreading is described as follows.
Selecting randomly a victim, the gossip about him or her is created at 
time $t=0$ by an originator which shares a bond with the victim.
At $t=1$ the originator only spreads the gossip to other 
nodes, which are connected to him-/herself and the victim.
The spread continues until all reachable acquaintances of the victim know 
it, as illustrated by the squares connected by dashed lines in 
Fig.~\ref{fig1} for a real friendship network~\cite{schools}.
Our dynamics is therefore like a burning
algorithm~\cite{burning}, starting at the originator but
limited to sites that are neighbors of the victim. 

To measure how effectively the gossip - or, in general, the information - 
attains the acquaintances of the
victim, we define the spreading factor as $f=n_f/k$, where
$n_f$ is the total number of people who eventually hear the gossip and
$k$ is the degree of the the victim.
In addition, we also define the spreading time $\tau$ which defines 
the minimum time it takes to reach this fraction $f$ of acquaintances,
giving a measure of how far these connected acquaintances are from 
each other.
It is important to note that $f$ and the standard definition of
clustering coefficient $C$~\cite{watts98,Amaral00} are different quantities,
since the later only measures the number of bonds between neighbors 
and contains no information about how such bonds distribute among
the victim's acquaintances.

We start in Sec.~\ref{sec:first} by studying how
such kind of information spreads in different networks, namely 
in scale-free and in small-world networks. Some analytical considerations
will be present for the particular case of the Apollonian 
network~\cite{hans05}.
The results of such artificial networks are also compared to
the ones obtained with an empirical network of social contacts
recently obtained from an U.S.~School survey~\cite{schools}, where 
friendship acquaintances were rigorously defined~\cite{schools,prl}.
There are also situations where the information about the target-node 
can be of interest 
beyond the first neighbors, like the case where the victim is a movie
star, yielding a scenario similar to the one of usual rumour propagation
or even epidemic spreading~\cite{telogama}.
These cases will be considered in Sec.~\ref{sec:higherlevels}.
Since the tendency for spreading information does not always implies
that its transmission will be certain, we introduce in 
Sec.~\ref{sec:q} a probability for each node to spread the information and 
study the main effects on the spreading dynamics.
Discussion and conclusions are given in Sec.~\ref{sec:conclusions}.
\begin{figure}[t]
\begin{center}
\includegraphics*[width=8.5cm]{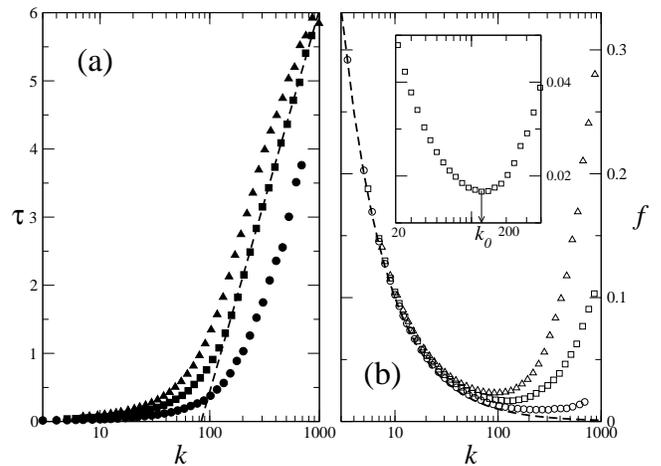}
\end{center}
\caption{\protect
  {\bf (a)} Spreading time $\tau$ in a Barab\'asi-Albert scale-free network 
  and 
  {\bf (b)} the spreading factor $f$, both as a function of $k$: 
  $m=3$ (circles), $m=5$ (squares) and $m=7$ (triangles). 
  The dashed line in (b) indicates $f=1/k$.
  The inset in (b) is a zoom of the plot for $m=5$ emphasizing the optimal 
  degree $k_0$ which minimizes the gossip spreading (see text).
  In all cases, $N=10^4$ nodes, averages over $500$ realizations are
  considered, and logarithmic binning in $k$ is used.}
\label{fig2}
\end{figure}

\section{Spreading information over first neighbors}
\label{sec:first}

We consider first a Barab\'asi-Albert (BA) scale-free network~\cite{barabasi99}:
starting with a small number $m$ of nodes fully connected to each other
one adds iteratively one new node with $m$ initial links attached to the
nodes of the network with a probability proportional to the node degree.

In Fig.~\ref{fig2}a we show the average spreading time $\tau$
as a function of the degree $k$ in a scale-free network with 
$N=10^4$ nodes and $m=3,5$ and $7$. 
In all cases, for large values of $k$, $\tau$ scales 
logarithmically with the degree
\begin{equation}
\tau = A + B\log{k}
\label{tau-k}
\end{equation}
where for this case $A=-10.77$ and $B=2.433$ defines the dashed line in 
Fig.~\ref{fig2}a.

For the same values of $m$ we plot in Fig.~\ref{fig2}b the dependence of 
the spread factor with the degree. Curiously, one sees an optimal 
degree $k_0$ for which the spreading factor attains a minimum (see inset).
This optimal value lies typically in the middle range of the degree spectrum
showing that the two extreme situations of having either few or many 
neighbors enhance the relative broadness of the information spreading.
Further, a closer look shows that for small degrees the 
values of $f$ coincide with $f=1/k$ (dashed line) while for larger degrees
$f$ deviates from $1/k$ with a deviation which increases with $m$.
Thus, while initially ($t=0$) the spread factor is always $f=1/k$ 
(dashed line), for the subsequent time-steps one observes that nodes with 
small degrees remain on average at $f=1/k$ while for large degrees the 
spread factor increases up to a maximal value. 
\begin{figure}[htb]
\begin{center}
\includegraphics*[width=8.5cm]{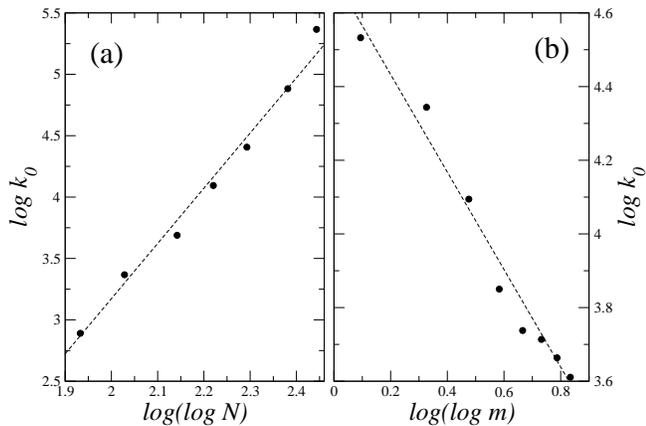}
\end{center}
\caption{\protect
  The optimal degree $k_0$ in a BA network as a function 
  {\bf (a)} of $N$ fixing $m=5$ initial outgoing connections and
  {\bf (b)} of $m$ for $N=10^4$ nodes.
  The average degree is $\langle k\rangle =2m$.
  The dotted lines have slopes of $a=4.64$ and $-b=-1.34$
  (see Eq.~(\ref{k0-N-m})).}
\label{fig3}
\end{figure}

The dependence of the optimal value $k_0$ on the two parameters
$N$ and $m$ is studied in Fig.~\ref{fig3}.
Here, we observe that the optimal degree $k_0$ yields approximately
\begin{equation}
k_0\propto \frac{(\log{N})^a}{(\log{m})^b} .
\label{k0-N-m}
\end{equation}
\begin{figure}[htb]
\begin{center}
\includegraphics*[width=4.3cm]{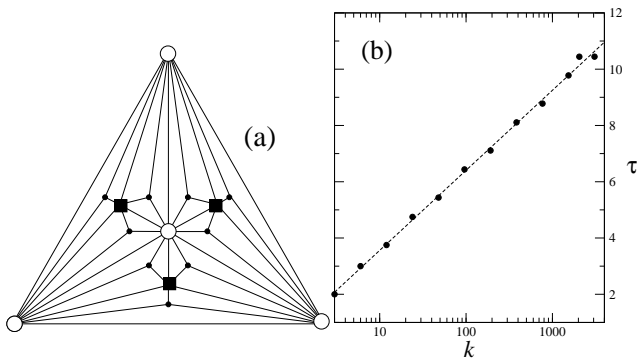}%
\includegraphics*[width=4.1cm]{fig04b_fofoca.eps}
\end{center}
\caption{\protect
  {\bf (a)} Illustration of the first three generations of an Apollonian
  network (see text).
  {\bf (b)} Spreading time $\tau$ for the spreading factor to 
  attain the maximal value $f=1$ where the dashed line can be obtained 
  analytically (see text) yielding an expression as in Eq.~(\ref{tau-k}) 
  with $A=-0.28$ and $B=1.1$.}
\label{fig4}
\end{figure}

The scale-free networks considered above are probabilistic. 
In other contexts, deterministic
scale-free networks have been proposed~\cite{hans05,dorogpseudo},
as a way to construct perfect hierarchical networks.
One of such networks is the Apollonian network.
The Apollonian network is constructed in a purely deterministic
way~\cite{hans05,pre04} as illustrated in Fig.~\ref{fig4}a:
one starts with three interconnected nodes, defining a triangle;
at $n=0$ (generation $0$) one inserts a new node at the center of the triangle 
and joins it to the three other nodes (white circles in Fig.~\ref{fig4}a), 
thus defining three new smaller triangles; at iteration $n=1$ one adds at 
the center of each of these three triangles a new node (squares), 
connected to the three vertices of the triangle, defining nine new triangles 
and then for generation $n=2$ one node (black circles) at the center of each 
of these nine triangles and henceforth.
The number of nodes and the number of connections are given respectively by
$N_n = \tfrac{1}{2}(3^{n+1}+5)$ and $L_n = \tfrac{3}{2}(3^{n+1}+1)$.
The distribution of connections obeys a power-law, since the number of
nodes with degree $k=3,3\cdot 2,3\cdot 
2^2,\dots, 3\cdot 2^{n-1},3\cdot 2^{n}$ and $2^{n+1}$ is equal to
$3^n,3^{n-1},3^{n-2},\dots,3^2,3,1$ and $3$, respectively.
Thus one has $P(k)\propto k^{-\gamma}$ with $\gamma=\ln{3}/\ln{2}$.

One main difference from the BA network is that, for
Apollonian networks $f=1$ independently of $k$, due to the hierarchical 
structure shown in Fig.~\ref{fig4}a.
In Fig.\ref{fig4}b one observes the logarithmic behavior of 
$\tau$ similar to the BA case.
In the Apollonian case the logarithmic behavior can even be derived 
analytically as follows.
From Fig.~\ref{fig4}a one sees that vertices belonging to the $n$th
generation communicate with each other through $n$ steps
thus $\tau\propto n$. Since the degree of the $n$th generation is
given by~\cite{hans05} $k=3\times 2^{n-1}$, one obtains the logarithmic
dependence of $\tau$ shown in Fig.~\ref{fig4}c, where the dashed line
yields the expression in Eq.~(\ref{tau-k}) with $A=-0.28$ and $B=1.1$.

Next, we show that the main results obtained for the
scale-free networks above are also characteristic
of real empirical social networks.
For that, we study the model for information propagation on a real 
social network, namely, the one extracted from empirical data 
obtained in an extensive 
study done within the National Longitudinal Study of Adolescent 
Health (AddHealth)~\cite{schools} at the Carolina Population 
Center.
The data comprehends a survey done between 1994 and 1995 in
$84$ American schools evaluating an in-school questionnaire to $90118$ 
students. The students are separated by the school they belong to 
and therefore there are $84$ networks with sizes ranging from 
$\sim 100$ to $\sim 2000$ students.
The aim is to allow social network researchers interested in general 
structural properties of friendship networks to study the structural 
and topological properties of social networks~\cite{bearman04}.
In previous studies~\cite{prl,physicaD}, it has been shown that 
the main properties
characterizing the underlying networks from these data can 
be easily reproduced with a mobile agent model.
\begin{figure}[t]
\begin{center}
\includegraphics*[width=8.5cm]{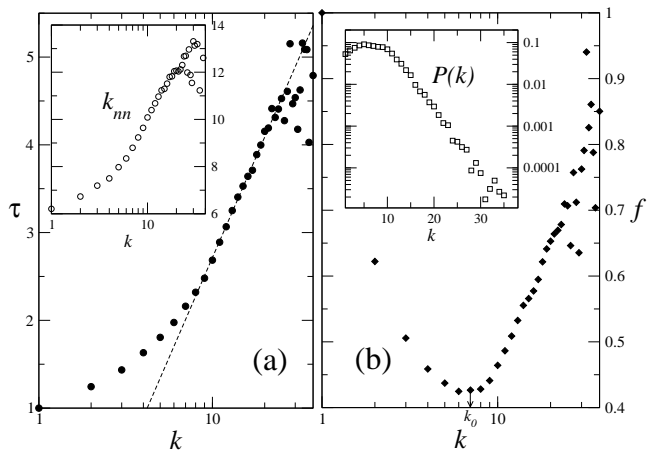}
\end{center}
\caption{\protect
        Propagation of information on a real friendship network of American  
        students~\cite{schools} averaged over 84 
        schools. In {\bf (a)} we show the spreading time
        $\tau$ as function of degree $k$, plotting in the inset, 
        the average degree $k_{nn}$ of neighbors of nodes 
        with degree $k$.
        In {\bf (b)} the spread factor $f$, both as a function of
        degree $k$, with the inset showing the degree distribution
        $P(k)$.}
\label{fig5}
\end{figure}

As shown in Fig.~\ref{fig5}a, while for small $k$ the spreading time 
grows linearly, for large $k$ it follows a logarithmic law given
by Eq.~(\ref{tau-k}) with $A=-2.84$ and $B=1.98$.
Here, the logarithmic growth of $\tau$ with $k$
follows the same dependence of the average degree 
$k_{nn}$ of the nearest neighbors~\cite{catanzaro}, as illustrated in 
the inset of Fig.~\ref{fig5}a.
Further, the non-trivial effect of having an optimal degree $k_0$ is
also observed in Fig.~\ref{fig5}b. 
For these schools one obtains $k_0\sim 7$ neighbors as an optimal
value for which $f\sim 0.42$, meaning that less than half of the
first neighbors are reached.
In other words, with less friends ($k<k_0$), 
the information is more able to reach a larger fraction
of them. But, contrary to intuition, the same occurs for the nodes 
having a larger number of friends.

Interestingly, information spreads in the same way either through
these empirical networks as on scale-free networks, although
the corresponding topological and statistical features are known to
be quite distinct~\cite{prl,physicaD}. For instance, as shown in
the inset of Fig.~\ref{fig5}b, the degree distribution $P(k)$ of the
school networks is typically exponential and not power-law.
Since the same optimal degree appears in BA networks,
one argues that the existence of this optimal number
is not necessarily related to the degree distribution of the network,
but rather to the degree correlations.
However, the relation between degree correlations, measured
by $k_{nn}$, and the logarithmic behavior of the spreading time is
not straightforward. While in the empirical network we find the same
distribution for both $k_{nn}$ and $\tau$, in BA and APL networks
$k_{nn}$ follows a power-law with $k$. 
In the case of uncorrelated networks, two and three-point correlations
reduce to simple expressions of the moments of the degree distribution.
Therefore, $f$ is independent of the degree, similarly to what is
observed for the density of particles as derived by Catanzaro et 
al~\cite{catanzaro2} in diffusion-annihilation processes on complex 
networks.
\begin{figure}[htb]
\begin{center}
\includegraphics*[width=8.5cm]{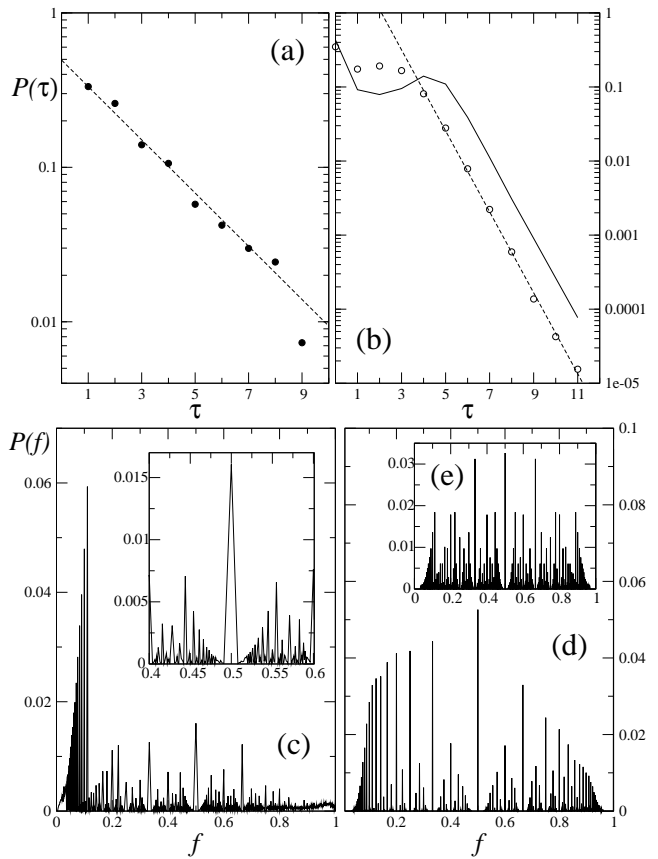}
\includegraphics*[width=8.5cm]{fig06cde_fofoca.eps}
\end{center}
\caption{\protect
        Distribution $P(\tau)$ of spreading times $\tau$ for 
        {\bf (a)} the Apollonian network of 8 generations, and 
        {\bf (b)} the real school network (circles) and the 
        BA network with $m=9$ and $N=1000$ 
        (solid line).
        The dashed lines indicate the best fit to the data for large
        $\tau$ values of Eq.~(\ref{distTauApoll}), with parameters
        $(1-\gamma)/B=-0.45$ and $-1.26$ for the APL in (a) and the real
        school network in (b), respectively.
        Below, the distributions of $f$ are shown for
        {\bf (c)} the BA network with the same 
        parameter values (inset magnifies the range $f\in [0.4,0.6]$),
        for
        {\bf (d)} the schools and for
        {\bf (e)} an artificial distribution of all possible
        fractions $f$ among the same number of nodes and neighbors.
        The highly positive skewness in $P(f)$ of both BA
        and schools networks are in strong deviation with the artificial 
        distribution, indicating a structure among the way 
        neighbors connect with each other (see text).}
\label{fig6}
\end{figure}

To go further with the characterization of information spreading on
networks, we next study the distributions, $P(\tau)$ and $P(f)$.
In Fig.~\ref{fig6}a we see that for the Apollonian network
the distribution $P(\tau)$ of the spreading time decays exponentially. 
This behavior can be understood if we consider that $P(\tau)d\tau=P(k)dk$ 
and use Eq.~(\ref{tau-k}) together with the degree distribution, 
$P(k) \propto k^{-\gamma}$, to obtain 
\begin{equation}
P(\tau) \propto \exp{\tfrac{\tau (1-\gamma)}{B}} ,
\label{distTauApoll}
\end{equation}
for large $k$. The slope in Fig.~\ref{fig6}a is precisely 
$(1-\gamma)/B = -0.17$ using $B=1.1$ from Fig.~\ref{fig4}c and 
$\gamma = 2.58$ from Ref.~\cite{hans05}.

For the school network $P(\tau)$ follows an
exponential decay for large $\tau$, as shown in Fig.~\ref{fig6}b,
and has a maximum for small $\tau$.
For comparison, we also plot in Fig.~\ref{fig6}b the distribution $P(\tau)$ 
for the BA network with $m=9$, which has a very 
similar shape but is shifted to the right, due to the larger minimal 
number of connections.
In both cases, the distribution is well fitted 
by an exponential.
The reason for the similiarities between empirical networks and 
BA networks at the particular value $m=9$ may be
related to the way the questionnaire was made at the schools:
each student should name their friends out of a maximal number of
$10$ acquaintances. From the similarities we could now argue that
in fact on average the students elected $9$ acquaintances each.

Figure \ref{fig6}c shows the distribution $P(f)$ for a scale-free 
BA network, while Fig.~\ref{fig6}d shows the same
distribution for the empirical networks. 
Before studying such distributions the following remarks should be 
taken into account. The spreading factor depends on the number $k$ of 
neighbors and consequently depends also on the network size, since 
the larger the network the larger the maximal number $k_{max}$ of 
neighbors a node may have. 
Furher, the spread factor varies always between the minimal value 
$0$ and the maximal value $1$ and for a given node with $k$ neighbors
the possible values are $f=0,1/k,2/k,\dots,(k-1)/k,1$.
Consequently, if for a specific network all the possible $f$-values 
appear with the same probability one should expect the distribution 
$P(f)$ to be symmetric around $f=1/2$ with discrete peaks at $n/k$ 
for $n=0,1,\dots,k$ and $k=1,\dots,k_{max}$. This artificial 
distribution is shown in Fig.~\ref{fig6}e, obtained from all
possible fractions constructed with all integers from 
$N=1$ to $1000$.

For BA networks, there is also a symmetry
in the vicinity of $f=1/2$ (Fig.~\ref{fig6}a).
However, different from an uniform 
distribution, one finds a strong asymmetry between small and 
large values of $f$: the most pronounced peaks are observed for 
$f\lesssim 0.1$.
This same behavior is observed for the empirical school networks, as shown
in Fig.~\ref{fig6}d, which is also strongly asymmetric when compared with
the corresponding uniform distribution of all possible values of $f$
sketched in Fig.~\ref{fig6}e.
The positive skewnesses indicate a higher frequency of low $f$-values 
than of larger ones, which indicates in fact that the neighbors of
nodes tend to form small separated sets of linked neighbors.
Consequently, one is able to address how the connections between
neighbors are groupped only by measuring the spreading factor for
the central node.
For the distribution $P(f)$ of the Apollonian network 
one trivially finds $P(f)=\delta (1-f)$ since the hierarchical structure
of the network always yields $f=1$, as mentioned before. 
\begin{figure}[t]
\begin{center}
\includegraphics*[width=8.5cm]{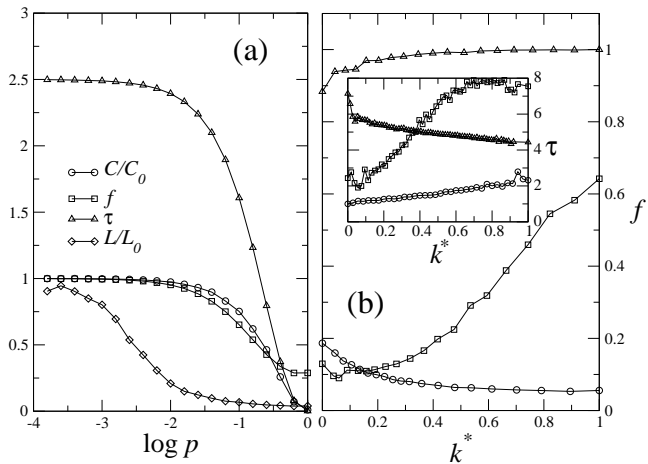}
\end{center}
\caption{\protect
         {\bf (a)} Propagation of information in small-world networks: 
         Spreading time $\tau$, clustering
         coefficient $C/C_0$ and spread factor $f$ as a function of
         the logarithm of the rewiring probability $p$ for the
         small-world lattice with $N=10^4$ sites. $C_0 = 1/2$ is
         the clustering coefficient of a regular lattice. In all cases
         we average over 100 configurations and $k_0=4$ (see text).
         {\bf (b)} Dependence of the spread factor $f$ on 
         $k^*=(k-k_{min})/(k_{max}-k_{min})$, for the random graph
         with $N=10^3$ sites and $p = 0.02$ (circles), $0.04$ (squares)
         and $0.08$ (triangles). In the inset: the spreading time $\tau$
         of the random networks for the same parameter values.}
\label{fig7}
\end{figure}

Social networks are usually small-world~\cite{strogatznat}, 
i.e., they are characterized by a high clustering coefficient
and a low average shortest path length.
Since we are interested in social systems we will next study the
propagation of information on artificial small-world networks, constructed 
as follows~\cite{strogatznat}.
One starts with a regular lattice where each node is attached to $k_0$ 
neighbors symmetrically displaced. Such regular network is characterized 
by a clustering coefficient $C_0$ and a shortest path length $L_0$.
In this regular network, all links are short-range.
Then, sweeping over all nodes one rewires with probability $p$ each link 
to a randomly chosen node. By doing this there will be on average 
$pk_0N/2$ long-range links. 

For $p=0$ the network is a regular structure where no long-range
links exist, yielding a large average path length and clustering coefficient.
For $p=1$ all links are long-range producing a random graph
structure where both average path length and clustering coefficient
are small.
Increasing $p$ from $0$ to $1$, one first observes
the decrease of the shortest path length $L$, when compared to $L_0$, 
and only for larger values of $p$ the decrease of the clustering 
coefficient $C$, as shown in Fig.~\ref{fig7}a.
Therefore, in the middle range between the decrease of $L$ and the decrease 
of $C$ one obtains the small-world effect where $L/L_0$ is small
and $C/C_0$ is large~\cite{barabasirev}.
As shown in Fig.~\ref{fig7}a this range is approximatelly
$-2\lesssim \log{p}\lesssim -1$.
In Fig.~\ref{fig7}a one also sees that both the spread factor $f$ 
starts to decrease at approximately the same value of $p$
as the normalized clustering coefficient $C/C_0$.

Figure \ref{fig7}b illustrates the variation of the spread factor 
as a function of the
degree in the particular case of a random network.
Instead of the above procedure with $p=1$ fixed, random networks can also 
be constructed by starting with $N$ nodes and introducing with probability 
$p^{\prime}$ one link between each pair of nodes.
Typically, in random networks there is a threshold $p^{\prime}_c$ beyond 
which different structure and dynamical features appear. This is also the 
case for gossip propagation. 
Figure \ref{fig7}b shows the behavior of $f$ in random networks for three 
illustrative values of $p^{\prime}=0.02, 0.04$ and
$0.08$, while the inset shows the corresponding spreading time. 
Since in random networks the average degree increases with $p^{\prime}$,
we choose to compute $f$ and $\tau$ as functions of 
$k^{\ast}=(k-k_{min})/(k_{max}-k_{min})$ in order to facilitate
comparison.
For $p^{\prime}=0.02$ and lower values both the spread factor and spreading 
time remain approximately constant, with $f\sim 1/k$ and $\tau\sim 1$.
Increasing the probability to $p^{\prime}=0.04$ increases the average degree 
per node and also the spread factor beyond its initial value $f=1/k$,
and consequently the corresponding spreading time, $\tau>1$, increases
with $k$.
Increasing even further the probability to $p^{\prime}=0.08$ and beyond,
more and more connections are introduced throughout the network, in 
particular among the neighbors of each node, which enables more 
nearest neighbors to know about the gossip.
Consequently, on average one obtains $f_{\hbox{max}}=1$ independently of 
$k$. This maximal value for such values of $p^{\prime}$ means that 
the spreading attains all the neighbors of the victim.
Therefore one should expect that the time to reach complete spreading
should decrease with $k^{\ast}$, which is what one observes in the inset 
of Fig.~\ref{fig7}b.

As a preliminary conclusion of this section one can state that,
although different in their structure, empirical social networks
behave similarly to scale-free networks when subject to propagation
of information over the first neighborhood of a particular
target-node.

\section{Beyond the first neighbors}
\label{sec:higherlevels}

In this Section we will study how $f$ and $\tau$ change
when the information is able to propagate beyond first neighbors.
For that, we consider two different regimes of information spreading.
In the first regime, it spreads among the first and 
second neighbors of the victim, and in the second 
it spreads throughout the entire network.
For the latter, there are two other quantities of interest that
we introduce here.
One is the total fraction $F_N$ of nodes who know and transmit the 
information, defined as
\begin{equation}
F_N=\frac{N_g}{N} ,
\label{FN}
\end{equation}
where $N_g$ is the maximal number of nodes in the entire network which 
already know the information and $N$ is the total number of nodes.
Second, the maximal spreading time $\tau_{max}$ defined as the number
of time-steps necessary to attain the fraction $F_N$.

\begin{figure}[t]
\begin{center}
\includegraphics*[width=8.5cm]{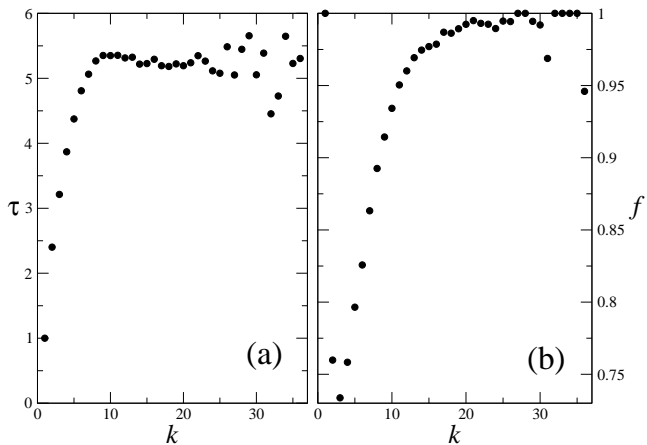}
\end{center}
\caption{\protect
  Information or gossip propagation through the first two neighborhoods
  in American schools:
  {\bf (a)} Spreading time $\tau$ as a function of $k$ and
  {\bf (b)} the spread factor $f$ as a function of $k$.
   As one sees the optimal number $k_0$ for which $f$ attains a minimum
   decreases significantly compared with the previous situation
   (see text).}
\label{fig8}
\end{figure}

Figure \ref{fig8} shows the spreading dynamics in the American schools
when it spreads among the two first neighborhoods of the victim.
The behavior is significantly different from the one observed
previously (compare with Fig.~\ref{fig5}).
From Fig.~\ref{fig8}a one sees that the spreading time becomes independent
on $k$ for large values deviating from the logarithmic dependence 
observed previously.

As for the spread factor $f$ shown in Fig.~\ref{fig8}b, one still
observes an optimal value minimizing the spreading of the gossip, but
this value is now much lower than the one found 
for propagation only among common neighbors of the originator 
and the victim.
Probably here, contrary to what happens in the previous case,
the optimal value vanishes when the network size or the number of
connections increase.
This conjecture will be reinforced next by studying
artificial scale-free networks.
\begin{figure}[htb]
\begin{center}
\includegraphics*[width=8.5cm]{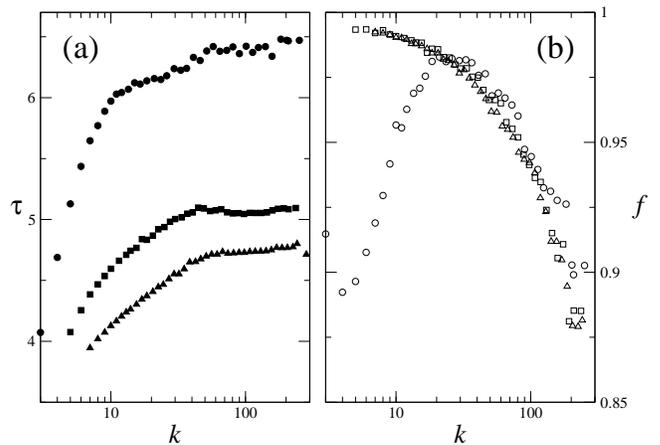}
\end{center}
\caption{\protect
  Propagation of information among first and second neighbors of a
  BA scale-free network.
  Here one sees 
  {\bf (a)} the spreading time $\tau$ as a function of the degree
  $k$ for $m=3$ (circles), $m=5$ (squares) and $m=7$ (triangles).
  {\bf (b)} Spread factor $f$ for the same 
  $m=3$ (circles), $m=5$ (squares) and $m=7$ (triangles).
  Here $N=10^4$, averages over $100$ realizations were considered
  and logarithmic binning in $k$ was used.}
\label{fig9}
\end{figure}

As illustrated in Fig.~\ref{fig9} the same behavior observed for the 
schools is also observed for BA networks.
Here, the results for
three different BA networks are shown for
$m=3$ (circles), $m=5$ (squares) and $m=7$ (triangles).
The spreading time $\tau$ attains also a constant value independent
on $k$ for large $k$-values (Fig.~\ref{fig9}a).
Obviously this plateau decreases with the minimal number $m$ of connections
and our simulations show that the dependence on $m$ is 
approximately logarithmic for small values of $k$.
This decrease happens 
because increasing $m$ increases the number of links per node, enabling
a faster propagation. Moreover the maximal value to which
$\tau$ converges for large $k$ can be explained as follows: since now 
the information spreads over first and
second neighbors, if the network has poor $k$-correlations, for sufficiently 
large $k$, all values of $k$ start to be present within the two first 
neighborhoods yielding an independence of $\tau$ on $k$.
The distribution of the spreading time presents also an approximatelly
exponential tail with a slope that increases with $m$.

As for the spread factor $f$, the optimal value $k_0$ is observed only
for small $m$ ($m=3$) and rapidly vanishes when $m$ is increased.
In fact, for large values of $m$ one finds large values of $f$ decreasing
with $k$ as $f\propto 1/k$.
This occurs independently of $m$.
Due to the large values of $f$, the distribution $P(f)$ has again a very 
pronounced peak at $f=1$.

While for these BA networks the results are quite different
when the two first neighbors are considered instead of only nearest 
neighbors, 
the Apollonian network displays an almost invariant behavior.
for an Apollonian network almost the same behavior remains.
The lack of sensibility to the increase of the neighborhood in 
Apollonian networks 
is a consequence of its hierarchical structure. 
Also for small-world and random networks similar results are obtained.
So, as preliminary conclusions one sees that in hierarchical networks and
in networks with small-world property it does not matter
if the information can be transmitted beyond the victim's acquaintances or
not: in one way or another everyone rapidly knows our secrets!

After seeing what happens in small neighborhoods, the
next question refers to the opposite limit, i.e., 
when all nodes are able to get the information from the originator.
Of course in this case the fraction $f$ almost always achieves eventually its
maximal value $f=1$, since the information eventually reaches everybody.
This is a similar situation of what happens with the spread of rumours
or epidemics.
Though, there is still the case when some neighbor of the victim has no
other friends and therefore the information cannot spread from or to it.
The main question now is not only to know the minimal time $\tau$
needed for the information to reach the maximal number of nearest neighbors of 
the victim, but also to compare it with the maximal time $\tau_{max}$ needed 
for the information to achieve the maximal fraction $F_N$ 
(see Eq.~(\ref{FN})) of nodes which are reached.
\begin{figure}[t]
\begin{center}
\includegraphics*[width=8.5cm]{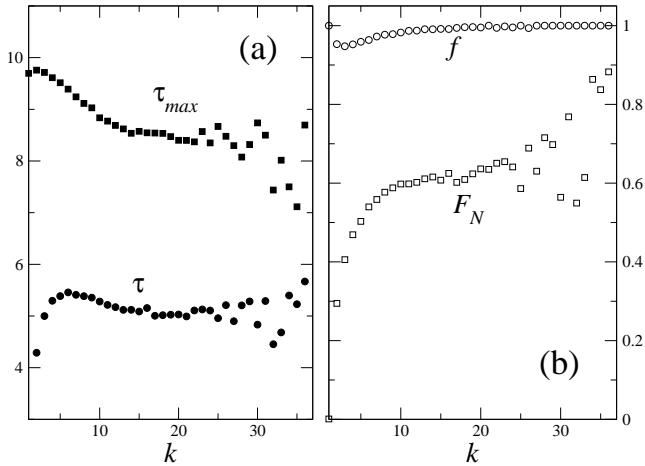}
\end{center}
\caption{\protect
  The spread of information through the entire school networks.
  {\bf (a)} Spreading time $\tau$ and maximal spreading time $\tau_{max}$
  as function of degree $k$. 
  {\bf (b)} Spread factor $f$ and total affected fraction $F_N$ as
  a function of $k$.} 
\label{fig10}
\end{figure}
\begin{figure}[htb]
\begin{center}
\includegraphics*[width=8.5cm]{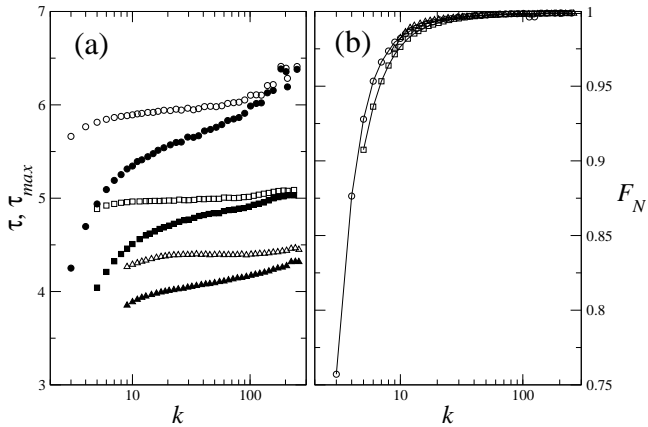}
\end{center}
\caption{\protect
  The propagation of information throughout an entire BA network.
  {\bf (a)} the spreading time $\tau$ and maximal spreading
  time $\tau_{max}$ as a function of the degree $k$ for $m=3$ (circles), 
  $m=5$ (squares) and $m=9$ (triangles).
  The total fraction $F_N$ of nodes that get the information is plotted in 
  {\bf (b)}. 
  In all cases, $f=1$ always (see text).
  Here $N=10^3$, averages over $100$ realizations were considered,
  and logarithmic binning in $k$ was used.}
\label{fig11}
\end{figure}

For the school networks, the behavior is illustrated in Fig.~\ref{fig10}.
From Fig.~\ref{fig10}a one sees that the behavior of $\tau$ is almost the 
same as in Fig.~\ref{fig8}a. The maximal time decreases with $k$ before 
attaining an approximatelly constant value. 
The large fluctuation for $k>25$ is due to poor statistics.
The decrease of $\tau_{max}$
for small $k$ occurs, since for victims
with less friends the successive neighborhoods through which the information
spreads comprehend a smaller amount of neighbors than when starting with 
a larger number of friends.

As explained above the spread factor is approximatelly one independently
of $n$, yielding a delta distribution $P(f)\sim \delta (1-f)$, while the 
maximal fraction $F_N$ increases fast for small $k$ and rapidly attains 
a more or less constant value around $F_N\sim 0.6$.
Therefore, no optimal number of friends is observed.

Figure \ref{fig11} shows what happens in the BA case.
As one sees from Fig.~\ref{fig11}a, both $\tau$ and $\tau_{max}$ decrease
with $m$. Further, for both quantities, $\tau$ (black symbols) and
$\tau_{max}$ (white symbols), a fast convergence to a logarithmic dependence
on $k$ is observed when $k$ increases. Interestingly, while the slope
as a function of $\log{k}$ differs between $\tau$ and $\tau_{max}$, in
each case it is approximately independent of $m$, being apparently
a feature of the scale-free topology.

In this situation one has always $f=1$.
As for $F_N$, very large values are now observed ($F_N > 0.7$) independently
of $k$ and $F_N$ increases very fast attaining $F_N\sim 1$ for $k>10$ 
neighbors (see Fig.~\ref{fig11}b). 
In other words, on BA networks, in order that all neighbors
of a certain victim get the information, it must spread throughout the entire 
network.
\begin{figure}[htb]
\begin{center}
\includegraphics*[width=8.5cm]{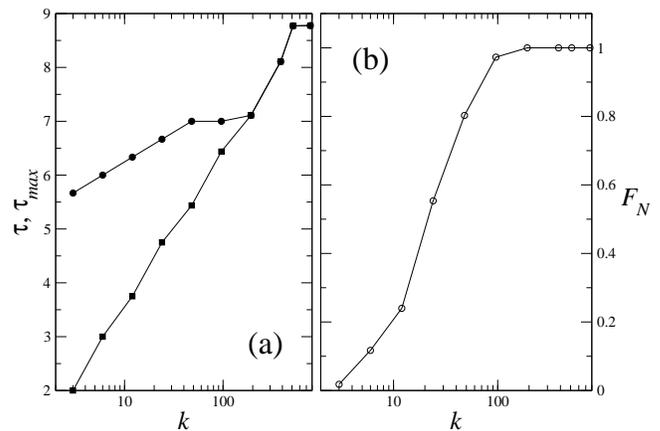}
\end{center}
\caption{\protect
  Propagation of information on an Apollonian network with $n=8$ generations:
  {\bf (a)} Minimal time $\tau$ and maximal time $\tau_{max}$
  and 
  {\bf (b)} the fraction $F_N$ between the total number of nodes
  which are reached by the information
  and the total number $N$ of nodes, both as functions of $k$.
  Here, $P(\tau)\propto P(\tau_{max})\propto P(k)\propto k^{-\gamma}$ 
  (see text).}
\label{fig12}
\end{figure}
\begin{figure}[t]
\begin{center}
\includegraphics*[width=8.5cm]{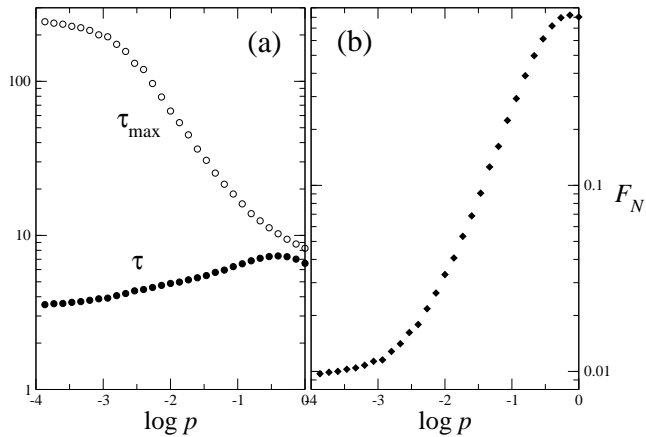}
\end{center}
\caption{\protect
  Information propagation in small-world networks when it spreads
  over the entire network.
  {\bf (a)} Spreading time $\tau$ and maximal time $\tau_{max}$ and
  {\bf (b)} total fraction $F_{N}$ as a function of the rewiring 
  probability $p$.
  Here $N=10^3$ nodes and averages over $100$ realizations were 
  considered.}
\label{fig13}
\end{figure}

Figure \ref{fig12} illustrates the case of the Apollonian network.
The value of $\tau_{max}\ge \tau$ increases more slowly 
with $k$, being both quantities equal for very large $k$ values. 
This similarity between both spreading times $\tau\sim\tau_{max}$
is in fact another evidence for the fact that in order to 
enable the information
to reach all neighbors it must spread throughout the entire network.
In fact, from Fig.~\ref{fig12}c one also sees that in the range where 
$\tau_{max}>\tau$, $F_N<1$, being equal to one only in the range
$\tau=\tau_{max}$.

Finally, we examine the case of small-world networks illustrated in
Fig.~\ref{fig13}.
From Fig.~\ref{fig13}a one sees that the spreading time $\tau$
increases almost linearly with the rewiring probability $p$ except
at the end for large values of $p$ (random network).
The maximal spreading time $\tau_{max}$ is very large for low rewiring
probabilities, due to a large average path length, and decreases
one order of magnitude in the range $-2<\log{p}<-1$ corresponding to 
small-world networks.
In fact, $\tau_{max}$ follows the dependence of the average
path length on $p$.

As for the total fraction $F_N$ illustrated in Fig.~\ref{fig13}b one finds
the opposite dependence on $p$ than the one found for $\tau_{max}$:
for low (large) values of $p$ one finds low (large) values of $F_N$,
and a pronounced increase is observed throughout the entire small-world
regime. 
To explain this behavior one must use both the average path length and
the clustering coefficient, 
$L/L_0$ and $C/C_0$ shown in Fig.~\ref{fig7}a.
For random networks ($p=1$) the total fraction attains $F_N=1$ very 
fast due to the very short average path length.
For small values of $p$, although regular networks have an
average path length that is larger than in random networks,
the spreading time needed to attain $F_N=1$ is now proportional to $L$.
In the small-world regime however, the average path length is small but
the way the neighbors are connected isolates in some few cases nodes 
from the information spreading process. 
So, although small-world networks have large cluster coefficients 
as in regular networks, the long-range connections change significantly the 
local topology of a given node-neighborhood.

\section{Introducing a transmission probability}
\label{sec:q}

In all the previous results each friend 
will surely spread the gossip further.
Fortunately people are on average not as nasty as that.
One should expect that only a certain fraction $q<1$ of our friends
are not worth to be trusted.
In this Section we address this more realistic situation.

Since we do not have any sociological information about the topological
features of the `good' friends we 
introduce $q$ as a probability that a node has to spread
the gossip. For the particular case $q=1$ one reduces to the situations
studied previously.

Two possible ways of propagation may then occur.
One concerns a scenario where friendships connections are related to
contacts between the nodes at a given instant. In this situation
a certain individual tries only once, with probability $q$, to 
spread the information to its friends. Therefore, if the gossip is not 
`accepted' once it will never be.
Another scenario is of course when the spread is tried repeatedly
at each time-step.
We will start with this latter scenario and end with the more pleasant one 
where gossip is only able to spread from the nodes which heard 
it most recently.
\begin{figure}[t]
\begin{center}
\includegraphics*[width=8.5cm]{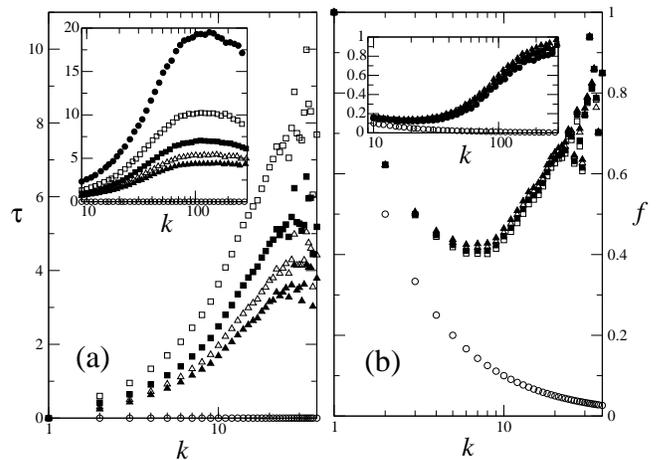}
\end{center}
\caption{\protect
        Information or gossip 
        propagation among first neighbors with probability $q$
        on a real friendship network of American students~\cite{schools} 
        averaged over 84 schools. In (a) we show the spreading time
        $\tau$ and in (b) the spread factor $f$, both as function of
        degree $k$. The insets show the same data for the BA
        network with $m=9$ and $N=1000$.
        At each time-step each node which knows the gossip tries to
        spread it.
        In all plots one has $q=0$ ($\circ$), $q=0.2$ ($\bullet$), $q=0.4$ 
        ($\square$), $q=0.6$ ($\blacksquare$), $q=0.8$ ($\triangle$) and 
        $q=1$ ($\blacktriangle$).}
\label{fig14}
\end{figure}

Introducing the new parameter $q$ in the model we go back to the 
first information spreading model studied in Section \ref{sec:first} where 
the gossip only spreads to friends of the victim.
At each time-step the neighbors which already know the gossip
repeatedly try to spread it to other friends of the victim.
Therefore, one expects to attain the same value of $f$ that one 
measured for $q=1$, but this time only after a larger spreading 
time, namely $\tau^{\prime}=\tau/q$.
Figure \ref{fig14} shows the result of such information propagation regime
for the school networks. and 
for several values of $q$.
The corresponding curves of $f$ are plotted in Fig.~\ref{fig14}b.
\begin{figure}[htb]
\begin{center}
\includegraphics*[width=8.5cm]{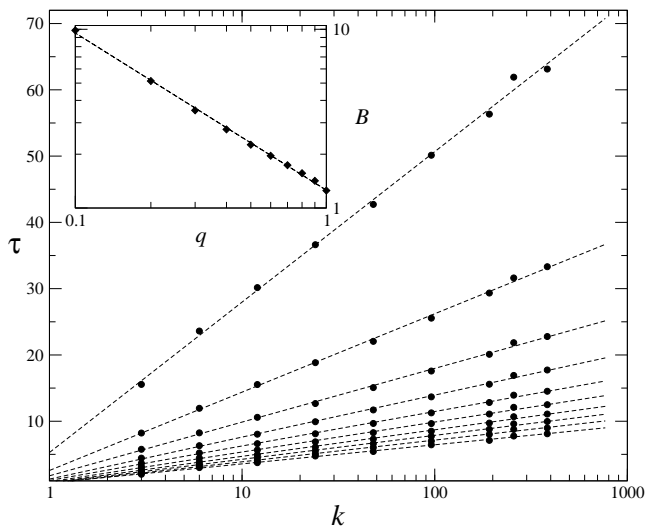}
\end{center}
\caption{\protect
  Gossip propagation in an Apollonian network with $n=8$ generations,
  for $10$ values of probability $q=0,0.1,0.2,\dots,0.9$ and $1$.
  The slope $B$ of the dashed lines which fit the data decreases
  with $q$ as shown in the inset where the line yields
  $B=\exp{(0.23-0.88\log{q})}\sim 1/q$ (see text).}
\label{fig15}
\end{figure}
\begin{figure}[t]
\begin{center}
\includegraphics*[width=8.5cm]{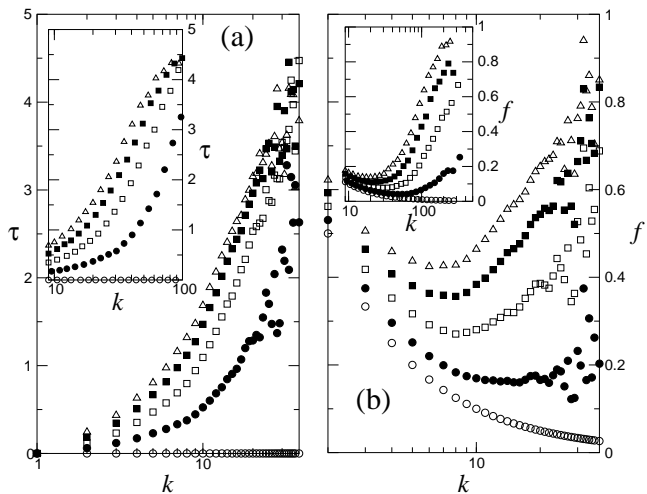}
\end{center}
\caption{\protect
        Propagation of gossip 
        among first neighbors on a real friendship 
        network of American students~\cite{schools},
        when persons to which a gossip did not spread at the 
        first attempt will never get it. 
        In (a) we show the spreading time
        $\tau$ and in (b) the spread factor $f$, both as a function of
        degree $k$. The insets show the same data for the BA
        network with $m=9$ and $N=1000$.
        After knowing the gossip each node tries to spread it only
        once (see text).
        In all plots one has $q=0$ ($\circ$), $q=0.25$ ($\bullet$), 
        $q=0.5$ ($\square$), $q=0.75$ ($\blacksquare$) and
        $q=1$ ($\triangle$).}
\label{fig16}
\end{figure}

Of course for $q=0$ the spreading time is always $\tau=0$ and
the spread factor equals $f=1/k$ since only the node starting the
gossip will know it.
As expected, for all other values the spread factor coincides with the
one for $q=1$, while the spreading time preserves its logarithmic 
dependence on $k$ for large degrees, and the exponent 
increases with $1/q$,as explained below.

In the insets of both plots in Fig.~\ref{fig14} we show for comparison
the spreading time $\tau$ and spread factor $f$ for a BA
network with $N=1000$ and $m=9$. 
A strong deviation from the logarithmic dependence of the spreading
time is observed, due to the high number of initial outgoing
connections ($m=9$). 

The logarithmic dependence of the spreading time can be more easily 
seen when studying the Apollonian network as shown in Fig.~\ref{fig15}.
Here we plot the spreading time for $10$ different values of $q$ and fit
all of them with a logarithmic function as the one in Eq.~(\ref{tau-k}).
The corresponding slope $B$ as a function of $q$ is plotted
in the inset of Fig.~\ref{fig15}
and follows closely a hyperbolic behavior,
$B\sim 1/q$.
Thus, Eq.~(\ref{tau-k}) can be written more generally as
\begin{equation}
\tau \propto \tfrac{1}{q}\log{k} .
\label{tau-k_gen}
\end{equation}

Finally, we can also assume that the person to which a gossip did not
spread at the first attempt, will never get it. 
In this way, the gossip is a quantity which percolates through
the system.

In Fig.~\ref{fig16} we 
see the behavior of $\tau$ and $f$ for different values of $q$ for the
school networks and in the inset for the BA network. 
When the spreading probability $q$ decreases, the minimum in $f$ first 
shifts to larger $k$ and finally disappears.
The asymptotic logarithmic law of $\tau$ for large $k$
remains for all probabilities $q$. As in previous cases,
the BA network has a similar behavior 
as the school friendships. The Apollonian network, 
however, behaves quite differently: $\tau$ first
increases with $q$ and then eventually falls off to zero so that
there exists a special value $q_{max} \approx 0.75$ for which the 
spreading time $\tau$ is maximized.

\section{Discussion and conclusions}
\label{sec:conclusions}

In this paper, we studied a general model of information spreading
suited for different kinds of social information.
In the usual case of rumour or opinion propagation the information
spreads throughout the network, and all nodes are equally capable of
transmiting the information to their neighbors.
Two measures were proposed to characterize the spreading of such model,
namely, the spreading factor measuring the accessible neighborhood
around each node which can be reached by the information spreading,
and the spreading time which computes the minimum time to reach
such neighborhood.

Further, we have shown that by computing these quantities for each
node the resulting distributions give additional insight to the
underlying network structure on which the spreading takes place.
More precisely, the magnitude of the skewness of the distribution of the 
spreading factor gives a measure of how difficult it is to access
one neighbor, starting from another one. 
For positive values of the skewness, most of the pairs of neighbors 
are connected by some path of connections, while for negative 
values of the skewness, neighbors are more likely groupped in 
separated connected pairs.

In the particular case that the information is about a certain 
target-node and thus is of interest to a restricted neighborhood
around it, one yields a minimal model to study gossip spreading.
Applying such a scheme to artifical and empirical networks, we
found that, although different in their statistical properties,
information on empirical social networks seems to spread similarly
to what is observed in scale-free networks.
In both cases, the spreading time shows a logarithmic 
dependence on the degree, indicating small-world effect within 
the nearest neighborhood of the nodes.
Further, from the computation of the spreading factor we observed 
that there is a non-trivial optimal number of friends which 
minimizes the danger of being gossipped that depends on the size of 
the network and on total number of acquaintances in it.
We also showed that this optimal value is characteristic of either
scale-free networks or real social networks, but is not observed
in small-world networks, rising the question of what network properties
may give rise to the emergence of such an optimal value.

However, when the information spreads beyond the nearest neighbors, 
in a similar way as for propagation of rumours and epidemics, this
optimal value disappears with the spreading factor rapidly
converging to $f=1$. Also the logarithmic dependence of the spreading 
time no longer holds in this case.

Since one person does not in general spread information to all its 
neighbors, neither at the same time nor with complete certainty, we 
also studied regimes of information propagation where the
spreading from one node to another occurs with some probability
$q$.

Due to their particular features and assumptions, our concepts 
and measures
to address the propagation of information in networks could be suited to other
situations. For instance, in the case of the Internet, some trojan
horses need to connect to a specific host to download some data in
order to become effective. For them the spread factor should be a good
measure to assess the vulnerability to the spreading of this virus
attack. In this situation probably an experimental test of the emergence
of the optimal degree found in the cases stated here could be easier
to be implemented.

\section*{Acknowledgements}

The authors profitted from discussions with Constantino Tsallis,
Marta C.~Gonz\'alez and Ana Nunes.
We thank the {\it Deutsche Forschungsgemeinschaft} and the Max Planck Prize
(Germany) and CAPES, CNPq and FUNCAP (Brazilian Agencies) for support.



\end{document}